# On the role of electromagnetic phenomena in some atmospheric processes

S. N. Artekha[1] and A.V. Belyan[1]

[1]{Space Research Institute, Moscow, Russia}

Correspondence to: S. N. Artekha (Sergey.Arteha@gmail.com)

**Abstract**

The crucial part of electromagnetic phenomena in many atmospheric processes is verified by systematized data. The multilayered charged system of clouds represents some dynamically equilibrium structure kept by the ionic and polarization forces. The estimates of acting forces are presented and it is demonstrated the necessity to take into account the plasma-like subsystems' effect on some atmospheric phenomena, including the formation and the maintenance of the structure and characteristics of their movement.

## 1   Introduction

Electromagnetic phenomena are present in almost all processes in the world. A striking example here is the electrical activity manifested within thunderstorms. The electrical activity also takes place during passage of tornados. There are other examples of hazardous atmospheric phenomena. The most impressive of which (in sizes) are hurricanes, typhoons or tropical cyclones (TC). The present paper is mainly devoted to study of these large-scale, vortex atmospheric phenomena, but not them only.

TC's nature and tornado's nature have been systematically studied for more than 170 years (Hare, 1837; Riehl, 1954; Khain and Sutyrin, 1983; Sharkov 1997). Plenty of time and financial resources are been spent for these purposes in many countries of the world. It is related with the extraordinary practical and theoretical importance of the problem. Despite the progress reached in this field of research, the exhaustive algorithmic theory still has not been constructed (see some criticism in Dobryshman, 1994; Artekha et al., 2003). As this takes place, the hydrodynamic theories (convection plus Coriolis force) have been basically developed. However, though short-term guessing of trajectories is now possible, these theories are still not capable of answering the key issues concerning the physical mechanisms of the origin and the intensification of TCs, maintenance of their stationary phase, manifestation of geographical, temporal, frequency and other asymmetries.

Possibly, due to excessive specialization in science, the specialists working in atmospheric sciences remain indifferent to a large complex of phenomena (namely, electromagnetic phenomena) observed in this area of research. At the same time, experimental observations indicate the presence of extended charged regions in the TC structure and the presence of strong electromagnetic regions in the TC and tornado regions (Black and Hallet, 1999; Vonnegut, 1960; Ziegler and MacGorman, 1994; Winn et al., 2000; Krasilnikov 1997; Chalmers, 1967; Williams, 1989; Marshall and Rust, 1995; Byrne et al., 1989). In all similar phenomena one can see the mechanism of generation of charged particles by friction (one of possible mechanisms). Even in a phenomenon of quite different nature, such as dust storms, the charges are intensively generated on dust particles at their friction of each other. Apparently, the friction of ice pieces each about other plays an important role in snow tornados and in tornados generated between the clouds. In some phenomena there exists an evident mechanism of ion formation: for example, the tornados sometimes arise over the powerful fires or over the eruption of volcanoes (in both cases the tornados are linked to a particular source of maintaining the ions and rapidly destroy as soon as the contact with a source is lost).

The preliminary estimates (Artekha et al., 2003; Arteha and Erokhin, 2004) have demonstrated an important role of electromagnetic forces in the inflow to the axis, in the levitation of particles, that is, in organization of ascending flows and in formation of the structure of clouds. In these articles it is discussed the idea about the influence of electromagnetic forces on large-scale atmospheric vortices. Though, actually, the conventional simple comparison of forces in physics can help in elucidating the character of processes on a linear section only. The problem can be reformulated in another manner. Hydrodynamic equations tacitly include the implementation of any pre-specified process (under appropriate initial and boundary conditions). How the probability of transition to the TC state, which is just of our interest, changes under the effect of forces, which accomplish such a transition "purposefully"? In addition, one should take into account that there always exist temporal and spatial scales, for which the approximations we have chosen (even strongest and most correct ones) are sometimes violated. The systematic trend of a state occurs on these small parts of phase path under an effect of "targeted" forces. Thus, in the general non-linear case the ignorance of "targeted" forces is lawful only in the case, if their integral effect during the given process' lifetime is small as compared to really observed manifestations of the studied process. Besides, even in the customary hydrodynamic description, the system development trend is, all the same, always specified "implicitly" (either via artificial specifying a large initial vorticity, or by specifying a desirable form of the required solution). However, if this key issue has already been specified "manually", one could ask, what we are searching for as a matter of fact (from physical viewpoint, but not meteorological)? It would be desirable to find just the physical mechanism, which is responsible for transition to TC and for maintaining its structure.

In any case, the work we have carried out does not claim for substitution of existing developed theories, but supplements them by studying one more side of these complicated atmospheric phenomena.

The processes in upper layers of atmosphere (stratosphere, ionosphere etc.) can have active effect on the typhoon genesis. The mutually opposite phenomenon should also be observed, namely, the responses of the given tropospheric processes in the ozonosphere (Gushchin and Sokolenko, 1985). Disclosure of the role of TC energy exchange with the open system – the upper atmosphere – demands further study.

The main objective of this paper is as follows:

– To pay researchers' attention to the facts (not included in the conventional developed directions) concerning the role of the Earth magnetic field and of electromagnetic processes in TC, the role of TC interaction with the open system (upper layers of the atmosphere);

– To systematize the experimental data and carry out estimations of all influencing factors.

## 2   Some key observational data

It is well known that the zone of the genesis of typhoons is located between the latitude of 30°N and the latitude of 30°S (except the near-equatorial region ±5°, since Coriolis force is small). The necessary conditions include the warm ocean (26.1 °C and above); in modern theory it is also needed low vertical wind shear and pre-existing disturbance.

Let us present the "crucial" data set, which forces one to think about a possible role of electromagnetic factors in the atmospheric phenomena under study (Artekha et al., 2003). First of all, one should mention here the geographical asymmetry of the typhoon genesis (Khain and Sutyrin, 1983; Dobryshman 1994). So, the number of TCs originated and developed in the northern hemisphere is twice greater, on the average (from 1.5 times to 4 times!), than it takes place in the southern hemisphere. So, separation into the western and eastern hemisphere is purely conventional approach, from the hydrodynamic viewpoint; but, nevertheless, the distinct typhoon genesis asymmetry is observed: the number of TCs originated in the eastern hemisphere is twice greater, than a similar number in the western hemisphere. All pairwise differences of ocean surfaces (at the zone of a typhoon genesis) for the mentioned hemispheres are much less than this value.

The binding to temperature conditions (26.1 °C and above) of the ocean surface cannot fully determine the physics of these phenomena. So, in the north the TCs are observed above the northern latitude of 35°, and this is not the case in the south (Dobryshman 1994). Since the mean time of TC existence is about one week, and the necessary temperature conditions exist in many parts of oceans during even much larger period of time, it is strange that TCs do not arise there at all.

The average sizes of TCs in the Atlantic and Pacific oceans are different (Baibakov and Martynov 1986; Merill 1984): Pacific-ocean TCs are slightly larger in size than Atlantic ones, but Atlantic TCs have slightly greater rate of rotation than rate of rotation for Pacific-ocean TCs (that anticorrelates with the geomagnetic field value in the TC origination area).

It seems absolutely unclear event that TCs are completely absent in the necessary zone of the ocean, close to the South America and close to Africa from the Atlantics side (see Fig. 1, which is publicly available from the Internet). Generally speaking, the TC generation area is located not simply in the above mentioned zone (as it should follow from purely hydrodynamic reasons), but, more likely, at the last zone intersection with the region located around the geomagnetic equator. Possibly, there exists some threshold on the vertical component of the geomagnetic field: $|B_z| \geq 2\times 10^{-5}$ T (see Fig. 2 from the handbook "Physical quantities", 1991). What is it: the random coincidence or the same factor that violates Earth system's axial symmetry in the atmospheric phenomena (possible, it is an additional necessary condition)? We remind that the magnetic axis does not coincide with Earth's axis of rotation, but is inclined to it by 11.5° and shifted by 1140 km from the Earth center towards the Pacific Ocean.

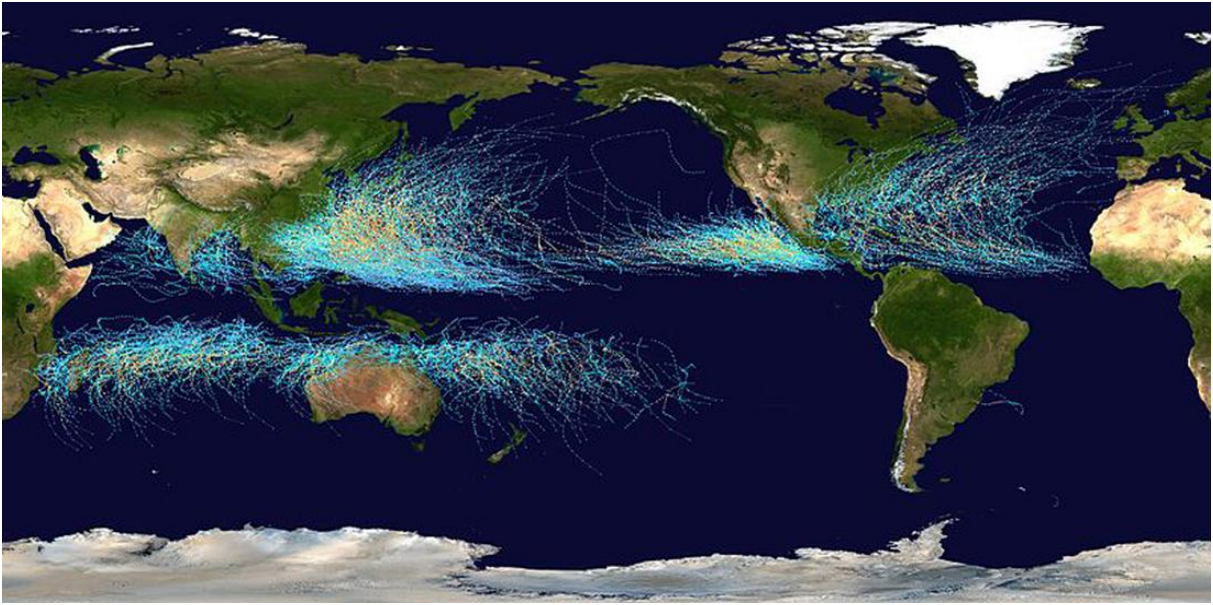

**Figure 1.** Global tropical cyclone tracks (from Wikipedia).

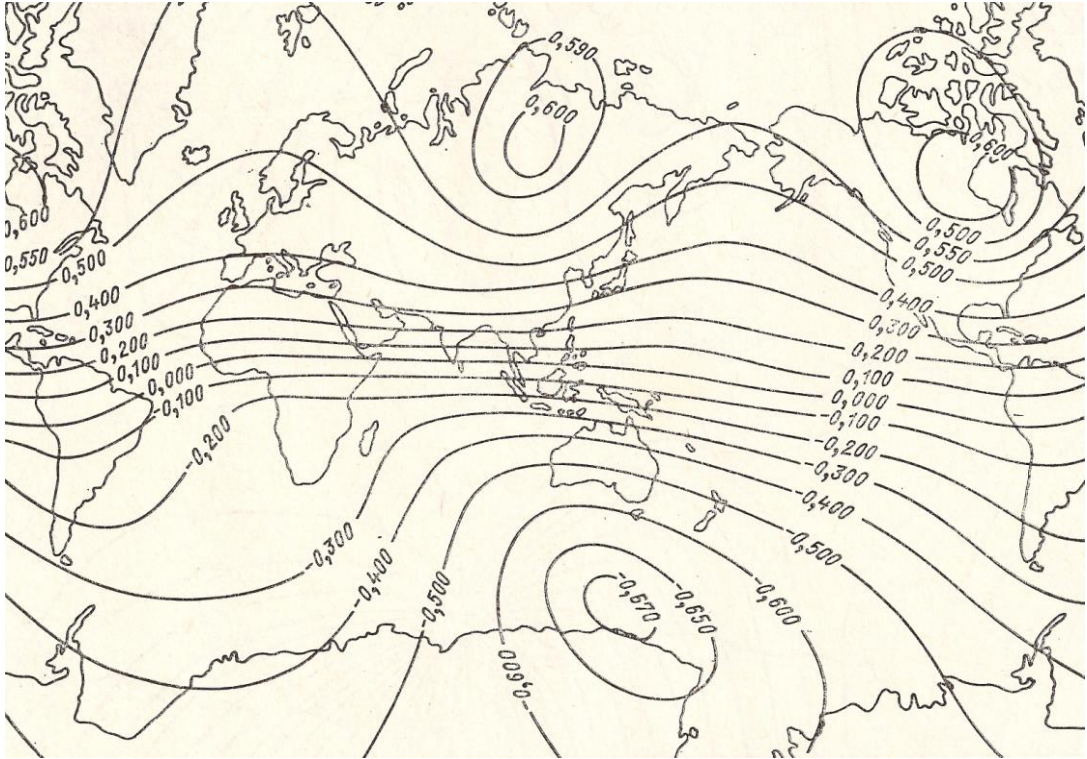

**Figure 2.** Vertical component of the earth's magnetic field (from the handbook "Physical quantities", 1991).

Many TCs arise at the middle of the trade-wind zone with an absolutely homogenous air mass. It means that the phrase about a large initial pulse and about temperature contrasts in the convergence zone is often unfounded.

The statement about the unique mechanism of converting motions into the vortex one via the contact with ocean is wrong: the TCs, even if they enter the land, do often exist for a long time, not to mention that a considerable part of TCs vanishes over the ocean. Besides, the rotation of opposite direction is observed over the TC (the anticyclone over the typhoon). This rotation obviously arises without contact with the ocean surface. For example, tornado's yoke drops from above; that is, the contact of aerodynamic flow with the ocean surface is absolutely unnecessary for originating and developing the vortex motion. TC does not also originate from the ocean surface, but it drops from some altitude.

If there was only purely hydrodynamic mechanism causing the increase of TC's moment of rotation, then the existing, rather large initial twists in both directions would be "picked up", and the TC rotating both clockwise and counter-clockwise should be observed often enough in both hemispheres (Coriolis force is not enough here to counter the beginning of this process). But this is not the case: the direction of TC rotation is fixed for each hemisphere (both northern and southern). This means that the additional mechanism can exist, which stands "above" the hydrodynamic and thermodynamic model and helps maintaining a strictly fixed structure of this phenomenon. Incidentally, in the northern hemisphere, the anti-cyclonic rotation of tornado is less common, and such tornadoes are more short-lived. But on these scales the effect of Coriolis force is practically negligible, therefore, the internal structure of a tornado (we assume – electromagnetic one) can also play an important role here.

The highest tangential velocity in TC is observed at certain altitude, and beginning at the other altitude another mechanism turns on, that causes anticyclonic rotation. Here, the fields of action of indicated mechanisms are close to the localization regions of charges with opposite signs. The jets of outflows from the top of TC are not axially symmetric, and their direction is not a random function. Possibly, these jets are highly influenced by charged particles, which tend to have drifts to the poles.

It would be tempting to consider all steady "rotational" atmospheric processes (cyclones, anticyclones, typhoons, dust-devils, tornadoes etc.) as having similar nature. All these phenomena arise, develop, then exist in the stationary phase for a long time (that is strange for aerodynamic phenomena) and move as a whole (intersecting terrains with a rather diverse relief).

Not only TCs can transfer into a conventional cyclone, but the opposite transition is also possible, when the subtropical cyclone transfers into TC. Apparently, all these processes can possess some additional common (non-hydrodynamic and non-thermodynamic) mechanism. The axis of a mid-latitudinal cyclone or an anticyclone is not vertical, as a rule, but it is highly inclined to the Earth surface. We remind that the Earth's magnetic field is also highly inclined to the surface, and the charged region tends to have the axis of rotation along the magnetic field. In reality, the inclination of the axis, precession and motion of a system as a whole are determined by several factors: the hydrodynamic rotating subsystem, connected below with Earth's surface and connected above with a corresponding flow, and the rotating charged subsystem, which tends to move according to EMHD-laws in self-consistent heterogeneous electric and magnetic fields. Two charged regions are simultaneously present in TC, and the axis occurs to be almost vertical, that can be related with electrical forces, which arrange oppositely charged rotating regions one under another and symmetrize the system. One could even observe reorientation of clouds' crystals in strong electrical regions (Saunders and Rimmer, 1999).

Conceivably, from this point of view it could be better understood the following: 1) why anticyclones are, as a rule, larger in size, than cyclones, but arise more seldom; 2) why in TC at the 12-km altitude the temperature increases by 10-12 degrees; 3) why the trajectories of TC motion often occur to be rather unpredictable; and many other issues.

If the purely hydrodynamic model was correct, then, apparently, the size of TC's eye would be uniquely related with the wind velocity. However, attempts to find such dependence have failed, that proves the existence of some disregarded parameters. Generally speaking, the size of TC's eye should not be related directly with TC intensity. Possible, one has to take into account, additionally, the quantitative and geometrical characteristics of all charged regions.

Thus, consideration of electromagnetic phenomena in TC and in other atmospheric phenomena can help to better elucidate all observations stated above. Here, the low-lying (4–8 km) negatively charged region can have influence on the cyclonic rotation, and the anti-cyclonic motion is supported by high-lying (10–16 km) positively charged region also.

If the hypothesis of electromagnetic nature of tornado and TC is correct, then some correlations should exist between the appearance of additional charged particles in Earth's atmosphere and origin of tropical depressions (Arteha and Erokhin, 2005). One more fact should be taken into account: in the equatorial zone the high-energy particles are observed; they cannot have cosmic origin (but, possibly, represent a consequence of some atmospheric processes).

The processes in upper layers of atmosphere (for example, the negative ionospheric disturbances arising during magnetic storms) can have active effect on the typhoon genesis. There exists some connection of troposphere and upper atmospheric layers (Sorokin et al., 2001; Vonnegut, 1997). The TC interaction with a larger open system (stratosphere, ionosphere, space) can be conventionally sub-divided into two aspects: 1) the indirect effect of the ozonosphere, ionosphere and cosmic factors on the processes occurring in TC (Reid, 2000; Rycroff et al., 2000; Tinsley, 2000; Stozhkov, 2003), and 2) the TC effect on the ozonosphere etc. (Rodgers et al., 1990; Stout and Rodgers, 1992; Nerushev, 1994; Nerushev, 1995; Nerushev et al., 1997; Sharma et al., 2004; Kazimirovsky et al., 2003). The above mentioned references support statements that will be made below.

We will begin with the first aspect. Possible effect of space factors (ionospheric disturbances) on the initiation of a thunderstorm genesis process, on number of typhoons and tornados and on some other physical atmospheric processes are studied insufficiently. Here several mechanisms can be in action, including those related with transformation of waves. At first, one should mention solar flares' influence on the origin of tropical depressions, including the direct effect of cosmic rays on the initiation of lightning flashes. Second, one should pay attention on the possibility of ionosphere interaction with plasma-like subsystems of a vortex via the electromagnetic waves. In addition, as the electric field changes, the plasma-like subsystems themselves can change by appearing some additional number of aerosols. The number of storms and lightnings correlates with the number of sunspots, with the geomagnetic index and with solar radio emission flow at wavelength of 10.7 cm, with the neutron flux, with the flux of cosmic rays. During flares the solar wind turbulence enlarges the flux of particles falling on the Earth. Cosmic rays have effect on the ionization (they modulate the production of ions) and on the atmosphere conductivity at altitudes from 3 to 35 km. The solar wind modulates the vertical current and influences the microphysics of clouds. The latter ones include the electric purification processes, albedo decrease and the change of passbands' width with respect to incident long-wave radiation, etc. One can propose some mechanism of changing passbands of the atmosphere. The ultraviolet radiation is related with ionization, production of ozone and free radicals. There exists also correlation between the cosmic ray flux and cloud cover via the additional ionization of droplets in clouds and nucleation. The correlation was found between the increase of electric field in the atmosphere and the growth of concentrations of aerosols. The influence of aerosols on weather and climate is a separate subject which is beyond the scope of this article (Budyko, 1974; Lee and Penner, 2010; Lohmann and Feichter, 2004). The ions, in their turn, influence atmospheric processes via: 1) charge-dependent chemical reactions, 2) forming of droplets and ice pieces during condensation on nuclei, 3) the currents in the global electric circuit (the maximum of charged particles' fluxes and ions production in the atmosphere is located at altitudes of 12 – 17 km). Of course, here we are dealing with a statistical effect only (i.e., some additional "favorable" conditions), but not with conditions which fully determine the specific atmospheric process (in addition, similar effects may occur with some delay).

Now we will turn to the second aspect – the responses of the given atmospheric processes in the ozonosphere and above. Intensive atmospheric vortices, such as thunderstorms with tornadoes and TCs, are capable to cause really recorded disturbances of geophysical fields, which can be used for the diagnostic and prognostic purposes. The effect of processes occurring in TCs can be transmitted by various mechanisms: by acoustic waves (first of all, by the infra sound), by inner gravity waves and by electromagnetic waves. There exists correlation between the amplitude of infrasonic pressure pulsations and intensification of tropical depressions (for example, in comparing with the process of establishing pressure profile in the TC). It is known that the long-living structural formations are observed over the intensive typhoon genesis areas. They can be discovered, for example, from the changes of ozone concentration (Gushchin and Sokolenko, 1985). Ozone is of interest for observation, since it serves as an indicator of electrical activity of the atmosphere (including corona discharges on cloud droplets) and can participate in the charge formation and separation processes (due to high electrochemical and photochemical activity). The TC in the process of its evolution causes variations of the total content of ozone and of its vertical distribution, up to stratospheric altitudes. For example, the total ozone content (in the northern hemisphere) has a negative anomaly for a larger part around the TC (except the left quadrant in the direction of motion that is occupied by the positive anomaly). The relative decrease of ozone concentration can reach some tens percents in the troposphere at the TC phase. Here we note some interesting feature: some percents heightened values of total ozone content took place on the northern periphery for tropical depressions, which have been subsequently developed in the TC (such phenomenon was not observed for decaying tropical depressions). Here we note that the maximum deviation of concentration for Pacific Ocean's TC is less than the same value for TC in the Atlantic Ocean. This correlates with corresponding values of the geomagnetic field (here the process of concentration change at transition through the tropical depression in the TC is slower in the Pacific Ocean as compared to the Atlantic). The total ozone content in the atmosphere is in a strict negative correlation with the number of TCs. Apparently, TC effects on the ozonosphere can be conventionally separated, in terms of mechanisms, into purely hydrodynamic ones, hydrodynamic effects related with electromagnetic forces and purely electromagnetic effects (via the fields or charged particles).

Wilson's hypothesis states that thunderstorms represent global electric circuit's generators. Thunderstorms (including ones in tornados and hurricanes), as generators, keep the mean ionosphere's potential at the level of the order of 250 kV relative to the Earth. The global electric circuit is modulated by the Sun effect (Kundt and Thuma, 1999). The correlation is observed between the currents of thunderstorms and the mean charge (field). All over the Earth one can observe universal diurnal variations of potential's gradient (the so-called "Carnegie curve"): it has the maximum at 1900 UTC, which can be related with the activity of thunderstorms (in tornados and hurricanes as well) via the global electric circuit. Note that at this time instant the Sun is located at the highest point (with heightened ion formation) over the meridian with the lowest geomagnetic field.

Over the storms, in the emission process, one can observe: increase of ionization in the ionospheric E-layer, appearance of sporadic F-layers, increase of electron temperature and concentration in the F-layer. The electrostatic fields of thunderstorms (tornados, hurricanes) are heating-up the electrons especially highly at night in the lower ionosphere. This effect is magnified for charges at high altitude and can modify the optical emission level, for example, via the chemical balance in the D-region. Ionospheric temperatures and ion densities change depending on the lightning activity: during thunderstorms one could observe (1.2–1.7)-fold increase of electron temperature; a similar increase of the ion temperature was (1.1–1.5)-fold. The possible reason of this is related to generation of ultra-short and very short electromagnetic waves. The multiple reflections on random heterogeneities can result in the non-linear conversion into plasma waves. There also occurs conversion into the high-hybrid plasma waves. In this case the electron density irregularities arise with various scales of heterogeneities. The trigger emission can occur in the region between the thunderstorm (including in a hurricane or tornado) and the ionosphere. The frequency of the lower hybrid resonance was also recorded on the linear sonograms of electric and magnetic fields. The lower hybrid waves can be stimulated by intensive whistlers in the presence of density gradients; and whistler-related LH waves from a linear mode are heating-up the epithermal ions. The efficiency of the incident electromagnetic field conversion into the LH resonance wave is about 30 %. Between the thunderstorm (or between appropriate area of a hurricane) and ionosphere one can also observe the durable emission in the earth-ionosphere waveguide at frequencies not depending on the altitude. Of great importance in the atmosphere-ionosphere exchange is Schuman's resonance (electromagnetic radiation from lightnings trapped into the earth-ionosphere waveguide). One also knows the formation of global, standing resonance waves at frequencies of 8, 14, 20, 26 … Hz and at subsequent harmonics. The Schuman resonance correlates with the global lightning activity, with the activity of sprites. The very-low frequency (VLF) electromagnetic waves are recorded at storms.

## 3  Electromagnetic characteristics of some atmospheric processes

Let us consider electrical characteristics of intensive large-scale vortices, including typical strengths of electric and magnetic fields, values of charges of clouds, densities of a volume uncompensated charge, size of charged droplets etc. Here one should distinguish the mean (or most typical) values from the local values of quantities (they can differ some orders of magnitude). We note that on charged cloud's boundary there exists, as a rule, a screening layer that complicates interpretation of the data obtained remotely. Some typical values of electrical parameters are collected in (Arteha and Erokhin, 2005; see also references therein).

Cloud clumps (clusters) have larger size at tropical latitudes as compared to moderate ones, and a rather high altitude of cloud base's disposition. Depending on the altitude and distance from cloud's center, the following types of clouds are observed in TCs: altostratus (As), altocumulus (Ac), stratus (St), stratocumulus (Sc), cumulus (Cu), cumulonimbus (Cb), nimbostratus (Ns). Some typical characteristics of clouds interesting to us are presented in Table 1. Here $\Delta z$ is the thickness, $E$ is the electric field strengths, $n_q$ is the volume density of charge, $q$ is the charge of particles. The typical charges of precipitation particles are of the order of $q \sim 3\times(10^{-13}-10^{-14})$ C in the near-earth layer. The charge (+) was located at the top of clouds and the charge (–) – at the bottom in 75 % of cases for the dipole electrical structure of clouds. The region of the main negative charge had thickness of the order of 2000 m; the main positive charge was located 1 – 7 km higher. In cumulonimbus clouds the small (+) charge was located at the bottom, the main (–) charge was above it, and the main (+) charge was already located above the latter one. In electrified clouds the heterogeneities of both charge and electrical potential gradient could be excited under an effect of various mechanisms, including floatability waves. The extreme values of cloud's volume charge density were observed in layers with thickness of the order of 100 – 200 m. Note that large particles increase their charge at collision with small ones. The minimum thickness of clouds, which generate thunderstorms, grows in the low latitudes. So, at the New Delhi latitude this thickness equals 9600 m. The thickness of overcooled water layers grows up to 8200 m. The increase of thickness (power) of clouds results in growing vertical flows and increasing cloud's charge $Q$. The presence of ascending and descending jets contributes in growing $Q$, and the turbulence and convection promote generation of heterogeneities of the field $E$, in which it can reach break-through values. The inverted dipole structure of a cloud (+ –) is also observed, where the main positive charge is located at the bottom of a cloud, and the main negative one – at the top. The distance between main charges of a dipole is 1.5 – 6 km.

**Table 1.** Characteristics of some clouds.

| Object | $\Delta z$, m | $E$, kV m$^{-1}$ | $n_q$, C m$^{-3}$ | $q$, C |
|---|---|---|---|---|
| As | 1000 - 1300 | 20 - 280 | $\leq 10^{-11}$ | $\leq 10^{-11}$ |
| Ns | 1000 - 2700 | 20 - 280 | $10^{-10} - 10^{-9}$ | $3 \times 10^{-12}$ |
| Sc | 400 - 500 | <10 | $(0.35 - 0.69) \times 10^{-11}$ | $1.8 \times 10^{-16}$ |
| St | 400 - 450 | <10 | $\leq 10^{-11}$ | $1.8 \times 10^{-16}$ |
| Cu | 1000 - 5000 | 20 - 280 | $10^{-10} - 10^{-9}$ | $3 \times 10^{-12} - 10^{-11}$ |
| Cb | 1000 - 15000 | 20 - 2000 | $3 \times 10^{-9} - 10^{-7}$ | $10^{-11} - 10^{-10}$ |
| TC | 8000 - 16000 | 10 - 300 | $10^{-8} - 5 \times 10^{-8}$ | $\leq 5 \times 10^{-11}$ |

Although for TC the mean volume charge density is 10 – 50 nC m$^{-3}$, one could also observe the mean densities up to 10 μC m$^{-3}$ (and locally – even higher values). For TC the charges of some small particles vary from –500 pC to +200 pC with the most typical values from –50 pC to +30 pC.

In the electric field $E$ the charged particles acquire velocity $v = b \cdot E$, where $b$ is mobility. Typical mobilities of ions are presented in Table 2. The ion velocity, at normalizing the quantities on typical values, is determined by the expression $v_i = 10(E/E_0)(b/b_0)$ m s$^{-1}$, where $E_0 = 10^5$ V m$^{-1}$, $b_0 = 10^{-4}$ m$^2$V$^{-1}$s$^{-1}$. Note that the contribution of free electrons to slow large-scale processes is unessential, because during the characteristic time of $\sim (10^{-7} - 10^{-2})$ s they stick to neutral molecules, forming negatively charged light ions of $H_3O+(H_2O)_6$ type. For aerosol ions the characteristic value of mobility is of the order of $2.3 \times 10^{-7}$ m$^2$V$^{-1}$s$^{-1}$. The typical size of aerosols is $(0.01 - 0.2)$ µm.

**Table 2.** Mobilities of ions.

| $r$, µm | $b$, m$^2$V$^{-1}$s$^{-2}$ |
|---|---|
| $6.6 \times 10^{-4}$ | $10^{-4}$ |
| $6.6 \times 10^{-4} - 8 \times 10^{-4}$ | $10^{-6} - 10^{-4}$ |
| $8 \times 10^{-4} - 2.5 \times 10^{-2}$ | $10^{-7} - 10^{-6}$ |
| $2.5 \times 10^{-2} - 5.7 \times 10^{-2}$ | $2.5 \times 10^{-8} - 10^{-7}$ |
| $> 0.057$ | $< 2.5 \times 10^{-8}$ |

For TC, on the average, the following structure of charged regions between the negatively charged Earth surface and the positive layer of tropopause is formed: near the Earth surface, at the TC center, there exists a small positive charge region; then, at the altitude of 4–8 kilometers, the most essential negative charge region is located; and, at last, at the altitude of 10–16 kilometers the positive charge region exists. Thicknesses of the last two charged layers are each about a kilometer. The described three-polar TC structure is observed for very intensive TCs. In reality, the number of charged regions in altitude often occurs to be greater than three. Here it was found that the greater is the number of charged regions, the lower is the vertical wind velocity, for example. Apparently, one can suppose that, as the TC evolves, the corresponding charged regions are combined, and their number decreases. Some TCs in their evolution have time enough to reach such an intensive phase; some TCs remain less intensive (and having a more complicated structure of regions and motions).

The multilayer electrical structures of a cloud cover are observed within thunderstorms (Marshall and Rust, 1995; Byrne et al., 1989). So, in the Wayene experiment the electrical structure of 8 charge layers was recorded, the layers being located bottom-up (up to 14 km) in the following order: + − + − + − + −. The thickness of layers varied from 400 m to 2600 m. Electric charge's volume density varied within the limits of $n_q \sim (0.1 - 2.4) \times 10^{-9}$ C m$^{-3}$, the maximum field $E_z \approx 115$ kV m$^{-1}$ was observed at the altitude $z \approx 6.3$ km. In the case of isolated super-cellular storm (the Ada experiment) maximum $E_z \approx 126$ kV m$^{-1}$ was observed at the altitude $z \approx 4.8$ km. The disposition of charged layers in the altitude ascending order was as follows: + − + 0 + 0 − 0 − + − beginning with $z \approx 3.2$ km and up to $z \approx 10.2$ km. Thickness of layers was ≤ 700 m, maximum $n_q \sim 1.34 \times 10^{-8}$ C m$^{-3}$. In the Elgin experiment 7 charged layers were recorded, which were located in the following order in altitude: + − + − 0 − 0 + − beginning with $z = 1.4$ km. The maximum of charge density $n_q \sim 0.8 \times 10^{-9}$ C m$^{-3}$ was at the altitude $z \approx 8$ km. The thickness of this layer was of the order of 2000 m. The extreme values of electric field $E_z$ in this experiment were: $E_z \approx 39$ kV m$^{-1}$ at the altitude $z \approx 3.8$ km, $E_z \approx -61$ kV m$^{-1}$ at the altitude $z \approx 7.8$ km, and $E_z \approx 48$ kV m$^{-1}$ at the altitude $z \approx 10.2$ km.

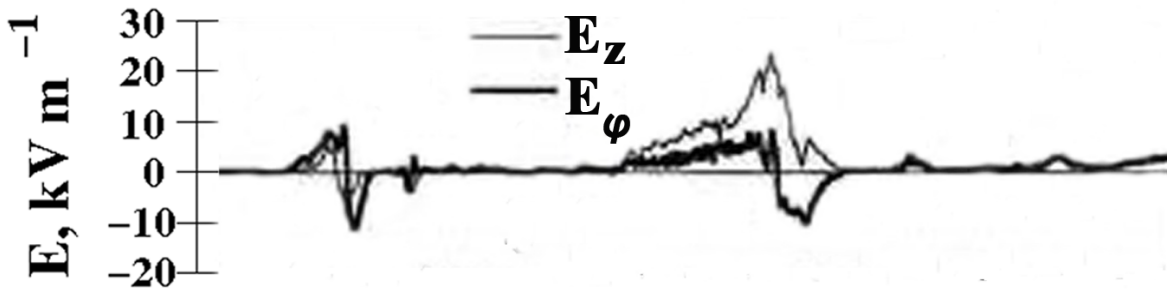

**Figure 3.** Electrical fields in tropical cyclones (from Black and Hallet, 1999).

The typical values of the electric field are tens of kV m$^{-1}$ in the central part of the TC (Black and Hallet, 1999). Depending on the altitude, they have either the same order of magnitude for tens of kilometers in radius, or vary significantly (near the hurricane's eye and rain bands - see, for example, Fig. 3). The typical behavior of the field during lightning intensification was as follows: at first, the field grows very slowly, and then suddenly reaches its peak value. When cloud's electric field reaches a break-through value, which is usually of the order of $10^6$ V m$^{-1}$ (but sometimes the breakthrough occurs at 400 kV m$^{-1}$), the discharge takes place with dropping the stored charge and releasing considerable energy. One average lightning stroke can heat up by 13 °C the air column of height 5 km and radius of 5 m. The pilot leader of diameter 10 cm causes formation of $10^{13} - 10^{15}$ pairs of ions per each centimeter of path. Frequently, lightning strokes pass through the same channel, in which the ions have not time enough to be recombined during the time interval between the strokes.

For TCs the rate of lightning strokes is quite variable (1 to 700) / hour and more. This value seems to be large, but taking into account TC's volume it can be small as it was noted by many researchers. Possibly, the electrical structure of TC is more regular, than in the case of mid-latitudinal thunderstorms, since it participates in maintaining the stationary phase of TC (the electrification of TCs is connected not only with lightnings!). However, the lightning flash rates can be considered as an indicator of the hurricane genesis (Leary and Ritchie, 2009; Price et al., 2009; Fierro et al., 2011).

In the developed typhoon the azimuthal motion of charged particles produces a rather high density of current's toroidal component $j_\varphi$. Normalizing the charge density and the azimuthal velocity for representative values, we obtain for a toroidal current's density component the value

$$j_\varphi = 5 \times 10^{-7} (n_q/n_0)(v_\varphi/v_{\varphi 0}) \text{ A m}^{-2},$$

where $n_0 = 10^{-8}$ C m$^{-3}$, $v_{\varphi 0} = 50$ m s$^{-1}$. For the azimuthal current flex cross-section $S_\varphi$ we have the following estimate of the large-scale azimuthal current:

$$I_\varphi = 250(n_q/n_0)(v_\varphi/v_{\varphi 0})(S_\varphi/S_{\varphi 0}) \text{ A},$$

where $S_{\varphi 0} = 500$ km$^2$. In its turn, the current $I_\varphi$ generates the poloidal magnetic field that compresses the current flex.

## 4  Estimation of forces and motions

Since the peak electromagnetic values are obtained for thunderstorms and tornadoes, and not for tropical cyclones, the subsequent evaluations will be done for TCs (here charged regions and electromagnetic fields are observed in the central area, mainly near eye's wall and rain bands). At first, we estimate the density of electric forces related with an surplus of charge of the same sign:

$$\mathbf{f}_q \sim n_q \mathbf{E}.$$

For eye's wall and charged TC regions (at the altitude near 6 km) this value lies within the range from $10^{-4}$ kg m$^{-2}$s$^{-2}$ up to 10 kg m$^{-2}$s$^{-2}$ with the mean value of $(10^{-3} - 10^{-2})$ kg m$^{-2}$s$^{-2}$. This is a rather large value. Let us compare it with the density of the force which retains the charged clouds:

$$\mathbf{f}_c = \frac{1}{2}(\varepsilon - 1)\varepsilon_0 \text{grad } \mathbf{E}^2.$$

This value for TC lies within the limits from $10^{-7}$ kg m$^{-2}$s$^{-2}$ to $6 \times 10^{-4}$ kg m$^{-2}$s$^{-2}$ with the mean value of the order of $10^{-4}$ kg m$^{-2}$s$^{-2}$, that is somewhat less, than the density of forces, which push away the charged particles in clouds. In order to explain the existence of charged cloud systems for such relation of forces, one must suppose some orderliness of charged particles in clouds. It is obvious that the total number of charged particles of both signs is some orders of magnitude higher than the number of charged particles, which provides the observed surplus of charge of some sign. As a result, the charged particles of various signs, on the average, alternate with each other, forming some semblance of an ordered structure kept as a whole by ponderomotive forces (ionic and polarization ones). Certainly, such an air-droplet (air-droplet-ice) "crystal" is dynamic one, which is present in the statistical (dynamic) equilibrium only. If we count the total sum of all forces (of attraction $+\sum_i q_0 q_1 / r_i^2$ and repulsion $-\sum_i q_0^2 / r_i^2$) between the chosen charged particle $q_0$ and all remaining particles, then we may be convinced that, due to such an orderliness, the system can be retained as a whole unit even at some surplus of charges. If we take into account that one should substitute the local quantities, rather than the mean value of field's square gradient, then the existence of a charged cloud occurs to be a norm, rather than an exception. At the excess of attractive forces the condensation increases, the droplets grow in size and fall down in the form of precipitation. So, the equilibrium is restored (for example, in the region of the largest change of the electric field – in rain bands – the dropping of precipitation occurs). So interesting is nature of this phenomenon.

Now we will estimate, what would be the velocity of a steady radial flow in TC in the case of balance between the friction force and the radial electric force with density *f*. For simplification we will take the linear flow in a planar layer with layer's height *h*. We will have the following expression for the maximum velocity of flow in a planar layer (Landau and Lifshitz, 1987):

$$v_{max} = -\frac{h^2 f}{12\eta},$$

where the dynamic viscosity of air is $\eta = 1.8 \times 10^{-5}$ Pa·s. Substituting the mean values that are typical for TCs, we will get for a one-kilometer layer the huge value: $v_{max} > 5 \times 10^6$ m s$^{-1}$. Given the fact that this layer is located at the altitude of 6 km, insignificantly improves the state of matter only. We will then consider the motion of a separate charged particle (if it would be alone) in air. In this case the steady uniform motion would have the following velocity:

$$v = \frac{N_e E}{6\pi a \eta},$$

where $N_e$ is the charge of the particle, $a$ is its radius. For example, we take the mean values for the central part of TC. Then, for the field strength of $3 \times 10^5$ V m$^{-1}$ the charged particles with $N_e = 20$ nC and with radius from 1.5 mm to 0.01 mm would acquire velocities from 20 m s$^{-1}$ up to 20 km s$^{-1}$! If, however, we take the micron-size crystal, charged by 1000 charges of electron, then the acquired velocity would be only 15 cm s$^{-1}$ relative to the medium (with its own flow rate). But, anyway, to find the flow rate of the medium itself, we should sum up the influence of all layer's charged particles on a medium. Schematically this can be done in the following manner. From (Landau and Lifshitz, 1987) we write the approximate formulas (in the spherical coordinate system), which express the components of flow velocities around a ball of radius $R$ in the far region ($r \gg R$):

$$v_r = u\cos\theta + \frac{3uR^2}{2\operatorname{Re} r^2}\left\{1 - \left[1 + \frac{\operatorname{Re} r}{2R}(1+\cos\theta)\right]\exp\left[\operatorname{Re} r(1-\cos\theta)/(2R)\right]\right\}$$

$$v_\theta = -u\sin\theta + \frac{3uR}{4r}\sin\theta \exp\left[\operatorname{Re} r(1-\cos\theta)/(2R)\right],$$

where $u$ is the flow velocity (directed along the Z axis), Re is Reynolds number. We pass into the Cartesian coordinate system (we are interested in the motion along the Z axis only) for the chosen particle, sum up in coordinates of all remaining ($N-1$) particles and get the correction – the coefficient to velocity (it only slightly differs from unity). Further, one should take into account that all influences are not independent, and each particle influences the «background basic velocity» for all remaining particles. In fact, this implies that to obtain a resulting correction, the earlier obtained coefficient should be raised into power ($N-1$). As a result, we will get again huge velocities amounted kilometers per second (this is natural – it is impossible to "bypass" the 1st Newton's law via the mean values). The only possibility to obtain reasonable (observable) quantities is as follows. The steady velocity of particles motion cannot be reached in principle! In reality, for example, not only a large number of negatively charged particles exists at the 6-km altitude, but a huge number of neutral particles and a large number of positively charged particles as well. The neutral particles move at the mean velocity approximately equal to the flow rate. The positive particles, whose number is great one and close to the number of negative particles, begin to move under electric field effect slightly faster, than the flux, and the negative particles begin to move in the opposite direction and have slightly lag to the mean flux motion. Each of these particles cannot gain essential velocity relative to the flux because of hindering (in addition to air's resistance) from collisions with neutral particles and opposite-sign particles, as a result of which the particles are pushed away (according to the momentum conservation law) to starting positions. One can schematically draw the elementary model of motion of such a "crystal element" (moving "ambipolarily on the average" toward the TC center) – the model of three particles in a medium – as follows on Fig. 4.

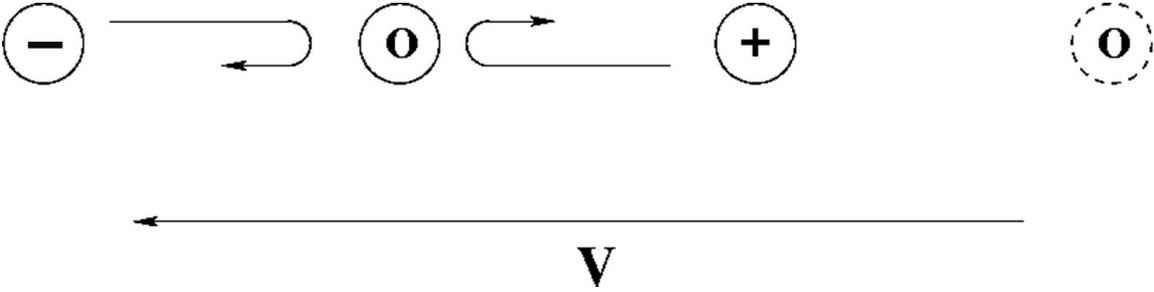

**Figure 4.** Model of three particles.

The particles spend most time at the positions shown here, because their relative velocity is zero at such positions (and at collision instants the relative velocities of particles are maximum and particles "jump" rapidly through such positions). In essence, this is the mechanism that converts the redundant work of electrical forces into the thermal energy and leads to rather low (really observed) velocities of radial motion. The micro-motions cannot be taken into account, in principle, in numerical calculations on mesoscales. Therefore one should take the quantities of the order of $2 \times 10^{-9}$ kg m$^{-2}$s$^{-2}$ or less (with corresponding signs) as an efficient value of electrical force density near the altitudes of 6 km and 12 km.

In addition to the radial motion, one should also consider the vertical motion. The motion of separate particles is influenced by the gravitational force, electrical force and aerodynamic drag force. First, we consider the balance of gravitational force and electrical one for separate charged particles of radius $r$, possessing charge $q$. Having taken $r = 0.1$ mm, $q = 10^{-11}$ C in the field with strength $E = 5 \times 10^4$ V m$^{-1}$ (values close to mean ones in the charged TC regions), we will have for a liquid drop: $k \equiv mg/(qE) \approx 8.2 \times 10^{-2}$, i.e. the effect of electrical forces exceeds the gravity force effect for these particles. As the liquid drop size decreases down to $r = 10$ microns, having taken the charge $q = 10^{-13}$ C, in the field $E = 5 \times 10^2$ V m$^{-1}$, we obtain $k \approx 4.3 \times 10^{-4}$. Let us consider now the balance of gravity force and aerodynamic drag force. In ascending air flows the uncharged drops of water can be retained at upper levels due to the aerodynamic drag force

$$F_R = \frac{1}{2}\pi r^2 C_R \rho_a v_z^2,$$

where $C_R$ is the aerodynamic coefficient, $\rho_a$ is the air density. The vertical velocity of an ascending flow, required for this purpose, is

$$v_z = \sqrt{\frac{8r\rho_w g}{3\rho_a C_R}}.$$

From here, for drops with radius $r = 0.5$ mm at the altitude $z = 5$ km for $T = -17$ °C and $\rho_a = 0.74$ kg m$^{-3}$, we obtain $C_R = 0.8$ и $v_z = 4.7$ m s$^{-1}$; for the drop radius $r = 0.25$ mm we have $C_R = 1.4$ and velocities $v_z = 2.5$ m s$^{-1}$ – such velocities are quite typical for TCs. Having changed the altitude $z = 10$ km, temperature $T = -50$ °C and density $\rho_a = 0.414$ kg m$^{-3}$, we obtain $C_R = 1.75$, $v_z = 3$ m s$^{-1}$. That it is quite real too. Concerning the balance of all three forces, one should say the following. The electric field strength reaches extreme values near the wall of TC eye, in the rain bands and close to the negatively charged TC region only. It is those places, where the electrical force plays considerable role (whereas in the remaining TC parts the dominant role in the balance of forces belongs to the gravity and aerodynamic drag forces). For example, the formation of a small low-lying positively charged region near the TC center is promoted by the electrical force acting (together with the gravity force) against the aerodynamic drag force. The electrical force essentially favors formation of a vast, negatively charged TC region at the altitude of about 6 km, because it helps particles to levitate (together with the aerodynamic drag force it reacts against the gravity force).

Consider now the TC rotation (the azimuthal motion is the key one). The plasma-in-the-magnetic-field model can serve as a useful model for describing this motion (This is only qualitative model for evaluation, but not for quantitative meteorological predictions). Let us recall the so-called L-H transition – spontaneous origination of rapid plasma rotation due to the $\mathbf{E} \times \mathbf{B}$- drift of charged particles (in the TC case we have the radial electric field and the vertical component of the magnetic field). It is essential that it is absolutely unnecessary to have fully ionized plasma, since the given mechanism will also work in the presence of some fraction of free charges in gas. The presence of neutral particles leads to the situation, where the momentum will transfer from charged particles into the rotational motion of gas as a whole at collision instants. As a result, the large-scale motion (rotation) will arise energetically due to reconstruction of inner energy of the whole system: not only the plasma subsystem, but also the whole adjacent region (the hydrodynamic system) starts rotating. All directions of rotation (both for the TC itself in both hemispheres and for the anticyclone over the TC) agree with observations. One can easily explain also the torus-like structure of TC with a calm region in typhoon's eye (Artekha et al., 2003). In the quiet atmosphere the number of ions is insufficient (the total number of ions does not exceed $10^9$ m$^{-3}$); just by this reason TCs arise rather rarely as compared to the other atmospheric phenomena (even the uncompensated surplus of charge in some regions of TC corresponds to the ion concentration of the order of $(10^{11} - 10^{13})$ m$^{-3}$, and the real number of ions is some orders of magnitude greater). If we estimate the magnetic force

$$\mathbf{f}_m = n \cdot [\mathbf{v} \times \mathbf{B}]$$

using the average concentration of a negative charge surplus in TC, then we would obtain an extremely small value of $\sim 2 \times 10^{-16}$ kg m$^{-2}$s$^{-2}$. But if the charge surplus in plasma-like subsystems constitutes a ten-millionth part of the total number of charged particles (what is quite possible), then such a force would already be capable to cause the observed macroscopic motion (for the central region of TC).

We consider a separate particle inside the charged region in the cylindrical coordinate system. Taking into account collisions with neutrals, the equations of forceless motion are:

$$eE_r(r,z) - m\nu_{in}(V_r - \overline{V_r(r,z)}) + eV_\phi B + mV_\phi^2/r = 0,$$

$$-m\nu_{in}(V_\phi - \overline{\Omega(r)}r) - eV_r B - \frac{mV_r V_\phi}{r} = 0.$$

Here the average velocities and the rate of neutral gas rotation are distinguished by a bar on the top, $r$ is the distance to the TC axis, $\nu_{in}$ is the rate of collisions with neutrals. As a result, for the azimuthal velocity we have the cubic equation:

$$V_\phi\left(m\nu + \frac{eE_r}{\nu r} + \frac{m\overline{V_r}}{r} + \frac{eB\omega_H}{\nu}\right) + V_\phi^2 \frac{2m\omega_H}{\nu r} + V_\phi^3 \frac{m}{\nu r^2} =$$

$$= m\nu\overline{\Omega}r - eE_r\frac{\omega_H}{\nu} - m\omega_H\overline{V_r}.$$

A charged particle (the negative ion with number $i$) acts on gas (due to collisions with neutrals) with the azimuthal force determined by the expression:

$$F_{\phi i} = m_i\nu_{in}(V_\phi - \overline{\Omega(r)}r).$$

Multiplying by the ion concentration and summing up over all sorts of ions, we will get the force acting on a unit volume of neutral gas from the side of a charged plasma-like subsystem:

$$\overline{F_\phi} = (V_\phi - \overline{\Omega(r)}r)\sum_i \nu_{in} m_i n_i(r,z).$$

To obtain the moment of forces acting on a system, one should multiply this expression by the corresponding radius and integrate over the whole volume of a charged subsystem:

$$M = \int_V r(V_\phi - \overline{\Omega(r)}r)\sum_i \nu_{in} m_i n_i(r,z)dV.$$

Due to the complicated process of microphysics and nonlinear relationships (the average value of a function is not equal to the function of the mean value) accurate estimates are very difficult. For example, choose the cloud layer of two-km thickness with radius of 100 km, the light ions with mass of the order of oxygen and make the estimations.

$\omega_H \sim -10^2$ s$^{-1}$, $\nu_{in} \sim 4\times10^9$ s$^{-1}$, $m \sim 6\times10^{-26}$ kg, $\Omega r \sim 50$ m s$^{-1}$,

$V_r \sim -10$ m s$^{-1}$, $E_r \sim -(10^4 - 10^5)$ V m$^{-1}$.

If the ion concentration will be supposed to be of the order of $5 \times 10^{12}$ m$^{-3}$ (the mean value for charged regions of TC), then, as a result, we will get for the moment of forces with magnetic spinning-up: $M \sim 4 \times 10^8$ kg m$^2$s$^{-2}$. The formula for the moment of gas' force of friction on the (disk) surface can be taken from (Landau and Lifshitz, 1987):

$$M \sim 0{,}97 R^4 \rho \sqrt{\nu \Omega^3}.$$

By making the substitution $\Omega^3 R^3 \to V_\phi^3$, as a result, one can obtain the following estimation for the value of a moment of the friction force (that determines the angular momentum outflow from a system):

$$M \sim 10^{11} \text{ kg m}^2\text{s}^{-2}.$$

If the total number of charges is at least two orders of magnitude greater than the number of uncompensated charges, then the moment will exceed the frictional torque: $M \sim 5 \times 10^{11}$ kg m$^2$s$^{-2}$, that is, in reality, the contribution of the electromagnetic forces in maintaining the rotation amounts from a few to several tens of percent for charged regions.

Of course, the electromagnetic processes do not provide the main contribution to the energy of tropical cyclones, but only help to maintain their structure. Apparently, the additional necessary condition for TC originating is accumulation of sufficient quantity of free charges in some atmospheric region. The ions act also as condensation centers, where the latent heat of evaporation is released. This is a triggering mechanism for converting huge stocks of thermal energy into the energy of motion. First of all, medium's inflow upwards arises, and air masses from an ambient space flow upwards to the TC axis on the place of ascending inflows. Some resemblance of a «cone in a trough» arises, which spins-up additionally, as air masses approach the axis due to angular momentum conservation.

The complexity of forecasting the TC motion consists additionally in the fact that, except the mass of neutral gas submitting to conventional equations of hydrodynamics, the charged rotating subsystem presents in TC. The value of resulting force and its direction can vary in an arbitrary manner depending on the value of charges of regions, currents and other TC characteristics, including TC coordinates. In all these motions the basic currents in TC will be convective ones from the motion of charged regions, rather than conductivity currents. The Earth magnetic field and existing electric fields vary along the TC trajectory. As a result, the given subsystems tend to move according to somewhat different laws, and for the TC sometimes we have the trajectory with hysteresis. Probably, that's why the loops and other unpredictable types of TC motion often arise, and the classic parabolic type of trajectory is observed in 47 percent of TCs only. If our suppositions are valid, the perspective can be opened not only to better forecast the origin, strengthening and motion of natural elements under study, but, possibly, even to control various phases of their evolution and trajectories of motion. Certainly, we do not imply here the "killing" of arisen TCs, because each phenomenon in the nature fulfills certain functions maintaining the total balance. In this case one can mention also the thermal and water balance, as well as the global terrestrial electric circuit. Of course, the nature will find the response mechanism on each unreasoned activity of a human being, but one should not permanently test our planet for strength.

## 5 Conclusions

Thus, in this article the key observational evidences are presented that electromagnetic phenomena play a significant role in many atmospheric processes. The multilayer charged system of clouds in TC is an analogue of a dynamic equilibrium ordered structure maintained by ionic and polarization forces. In the paper estimations of forces and mechanisms are made, and it is proved that the motion of plasma-like subsystems must be taken into consideration for more complete description of processes within thunderstorms, tornados and TCs. Electromagnetic forces are involved in generation and maintenance of the charged structure (including the distinct separation of movements inside TC) and can influence the motion of TC as a whole.